\documentclass[preprint,5p,twocolumn,times]{elsarticle}

\usepackage{amssymb}

\usepackage{graphicx}
\usepackage{epstopdf}
\usepackage{algorithmic}
\usepackage{algorithm}
\usepackage{array}
\usepackage{caption}
\usepackage[cmex10]{amsmath}
\usepackage{tabularx}

\providecommand{\norm}[1]{\lVert#1\rVert}

\journal{South African Journal of Science}

\begin{document}

\begin{frontmatter}


\title{High-speed detection of emergent market clustering via an unsupervised parallel genetic algorithm}


\author{Dieter Hendricks$^\ast$}
\author{Tim Gebbie}
\author{Diane Wilcox}

\address{School of Computational and Applied Mathematics, University of the Witwatersrand\\
					Johannesburg, WITS 2050, South Africa\\
					$^\ast$e-mail: dieter.hendricks@students.wits.ac.za
		}

\begin{abstract}
We implement a master-slave parallel genetic algorithm (PGA) with a bespoke log-likelihood fitness function to identify emergent clusters within price evolutions. We use graphics processing units (GPUs) to implement a PGA and visualise the results using disjoint minimal spanning trees (MSTs). We demonstrate that our GPU PGA, implemented on a commercially available general purpose GPU, is able to recover stock clusters in sub-second speed, based on a subset of stocks in the South African market. This represents a pragmatic choice for low-cost, scalable parallel computing and is significantly faster than a prototype serial implementation in an optimised C-based fourth-generation programming language, although the results are not directly comparable due to compiler differences. Combined with fast online intraday correlation matrix estimation from high frequency data for cluster identification, the proposed implementation offers cost-effective, near-real-time risk assessment for financial practitioners.
\end{abstract}


\end{frontmatter}


\section{Introduction}
Advances in technology underpinning multiple domains have increased the capacity to generate and store data and metadata relating to domain processes. The field of data science is continuously evolving to meet the challenge of gleaning insights from these large data sets, with extensive research in exact algorithms, heuristics and meta-heuristics for solving combinatorial optimisation problems. The primary advantage of using exact methods is the guarantee of finding the global optimum for the problem. However, a disadvantage when solving complex (NP-hard) problems is the exponential growth of the execution time proportional to the problem instance size \cite{luque}. Heuristics tend to be efficient, but solution quality cannot be guaranteed and techniques are often not versatile \cite{CDMMRT1996}. Meta-heuristics attempt to consolidate these two approaches and deliver an acceptable solution in a reasonable time frame. A large number of meta-heuristics designed for solving complex problems exist in the literature and the genetic algorithm (GA) has emerged as a prominent technique, using intensive global search heuristics that explore a search space intelligently to solve optimisation problems. 

Although the algorithms must traverse large spaces, the computationally intensive calculations can be performed independently. Compute Unified Device Architecture (CUDA) is Nvidia’s parallel computing platform which is well suited to many computational tasks, particularly where data parallelism is possible. Implementing a GA to perform cluster analysis on vast data sets using this platform allows one to mine though the data relatively quickly and at a fraction of the cost of large data centres or computational grids.

A number of authors have considered parallel architectures to accelerate GAs (see \cite{tirumalai,dessel,jaimes,pospichal,robilliard,brecheisen,bohm,kromer} as examples). While the work of \cite{kromer} is conceptually similar to the implementation proposed in this paper, a key difference is our choice of fitness function for the clustering scheme.

Giada and Marsili propose an unsupervised, parameter-free approach to finding data clusters, based on the maximum likelihood principle \cite{giada2002}. They derive a log-likelihood function, where a given cluster configuration can be assessed to determine whether it represents the inherent structure for the dataset: cluster configurations which approach the maximum log-likelihood are better representatives of the data structure. This log-likelihood function is thus a natural candidate for the fitness function in a GA implementation, where the population continually evolves to produce a cluster configuration which maximises the log-likelihood. The optimal number of clusters is a free parameter, unlike in traditional techniques where the number of clusters needs to be specified a priori. While unsupervised approaches have been considered (see \cite{omran} and references therein), the advantage of the Giada and Marsili approach is that it has a natural interpretation for clustering in the application domain explored here.

Monitoring intraday clustering of financial instruments allows one to better understand market characteristics and systemic risks. While genetic algorithms provide a versatile methodology for identifying such clusters, serial implementations are computationally intensive and can take a long time to converge to a best approximation. In this paper, we introduce a maintainable and scalable master-slave parallel genetic algorithm (PGA) framework for unsupervised cluster analysis on the CUDA platform, which is able to detect clusters using the Giada and Marsili likelihood function. By applying the proposed cluster analysis approach and examining the clustering behaviour of financial instruments, this offers a unique perspective to monitoring the intraday characteristics of the stock market and the detection of structural changes in near-real-time. The novel implementation presented in this paper builds on the contribution of Cieslakiewicz \cite{cieslakiewicz}. While this paper provides an overview and specific use-case for the algorithm, the authors are investigating aspects of adjoint parameter tuning, performance scalability and the impact on solution quality for varying stock universe sizes and cluster types.

This paper proceeds as follows: Section 2 introduces cluster analysis, focusing on the maximum likelihood approach proposed by Giada and Marsili \cite{giada2001}. Section 3 discusses the master-slave PGA. Section 4 discusses the CUDA computational platform and our specific implementation. Section 5 discusses data and results from this analysis, before concluding in Section 6. 

\vspace{-2mm}
\section{Cluster Analysis}
Cluster analysis groups objects according to metadata describing the objects or their associations \cite{everitt}. The goal is to ensure that objects within a group exhibit similar characteristics and are unrelated to objects in other groups. The greater the homogeneity within a group, and the greater the heterogeneity between groups, the more pronounced the clustering. In order to isolate clusters of similar objects, one needs to utilise a data clustering approach that will recover inherent structures efficiently. 

%
%

\subsection{The correlation measure of similarity}
The correlation measure is an approach to standardise the data by using the statistical interdependence between data points. The correlation indicates the direction (positive or negative) and the degree or strength of the relationship between two data points. The most common correlation coefficient which measures the relationship between data points is the \textit{Pearson correlation coefficient}, which is sensitive only to a linear relationship between them. The Pearson correlation is +1 in the case of a perfect positive linear relationship and -1 in the case of a perfect negative linear relationship and some value between -1 and +1 in all other cases, with values close to 0 signalling negligible interdependence.


\subsection{Clustering procedures}
Any specific clustering procedure entails optimising some kind of criterion, such as minimising the within-cluster variance or maximising the distance between the objects or clusters. 

\subsubsection{Cluster analysis based on the maximum likelihood principle}
Maximum likelihood estimation is a method of estimating the parameters of a statistical model. Data clustering on the other hand deals with the problem of classifying or categorising a set of $N$ objects or clusters, so that the objects within a group or cluster are more similar than objects belonging to different groups. If each object is identified by $D$ measurements, then an object can be represented as a tuple,  $\bar{x_i} = (x^{(1)}_i,..., x^{(n)}_i)$, $i = 1,...,N$ in a $D$-dimensional space. Data clustering will try to identify clusters as more densely populated regions in this vector space. Thus, a configuration of clusters is represented by a set $\mathcal{S} = \lbrace s_i,...,s_N\rbrace$ of integer labels, where $s_i$ denotes the cluster that object $i$ belongs to and $N$ is the number of objects \cite{giada2002} (if $s_i=s_j=s$, then object $i$ and object $j$ reside in the same cluster), and if $s_i$ takes on values from $1$ to $M$ and $M = N$, then each cluster is a \textit{singleton} cluster constituting one object only.

\subsubsection{Analogy to the Potts model}
One can apply super-paramagnetic ordering of a $q$-state Potts model directly for cluster identification \cite{BWD1996}. In a market Potts model, each stock can take on $q$-states and each state can be represented by a cluster of similar stocks \cite{BWD1996, KKM2000,giada2001}. Cluster membership is indicative of some commonality among the cluster members. Each stock has a component of its dynamics as a function of the state it is in and a component of its dynamics influenced by stock specific noise. In addition, there may be global couplings that influence all the stocks, i.e. the external field that represents a market mode.

In the super-paramagnetic clustering approach, the cost function can be considered as a Hamiltonian whose low energy states correspond to cluster configurations that are most compatible with the data sample. Structures are then identified with configurations ${\cal S} = \{ {s_i} \}_{i=1}^N$ for the cluster indices $s_i$, which represents cluster to which the $i$-th object belongs. This allows one to interpret $s_i$ as a Potts spin in the Potts model Hamiltonian with $J_{ij}$ decreasing with the distance between objects \cite{BWD1996, KKM2000}. The Hamiltonian takes on the form:
\begin{eqnarray}
H_g = - \sum_{s_i,s_j \in S} J_{ij} \delta(s_i,s_j) - \frac{1}{\beta} \sum_i h_i^{_M} s_i,
\end{eqnarray}
where the spins $s_i$ can take on $q$-states and the external magnetic fields are given by $h_i^{_M}$. The first term represents common internal influences and the second term represents external influences. We ignore the second term when fitting data, as we include shared factors directly in later sections when we discuss information and risk and the influence of these on price changes.

In the Potts model approach one can think of the coupling parameters $J_{ij}$ as being a function of the correlation coefficient \cite{KKM2000,giada2001}. This is used to specify a distance function that is decreasing with distance between objects. If all the spins are related in this way then each pair of spins is connect by some non-vanishing coupling $J_{ij}=J_{ij}(c_{ij})$. In this model, the case where there is only one cluster can be thought of as a ground state. As the system becomes more excited, it could break up into additional clusters and each cluster would have specific Potts magnetisations, even though nett magnetisation may remain zero for the complete system. Generically, the correlation would then be both a function of time and temperature in order to encode both the evolution of clusters, as well as the hierarchy of clusters as a function of temperature. In the basic approach, one is looking for the lowest energy state that fits the data. In order to parameterise the model efficiently one can choose to make the Noh ansatz \cite{Noh2000} and use this to develop a maximum-likelihood approach \cite{giada2001} rather than explicitly solving the Potts Hamiltonian numerically \cite{BWD1996, KKM2000}.

\subsubsection{Giada and Marsili clustering technique}
%
Following Giada and Marsili \cite{giada2001}, we assume that price increments evolve under Noh \cite{Noh2000} model dynamics, whereby objects belonging to the same cluster should share a common component: 
\begin{equation}
	\bar{x}_i = g_{s_i} \bar{\eta}_{s_i} + \sqrt{1 - g^2_{s_i}} \bar{\epsilon}_i .
\end{equation}
Here, $\bar{x}_i$ represents the features of object $i$ and $s_i$ is the label of the cluster that the object belongs to. The data has been normalised to have zero mean and unit variance. 
$\bar{\epsilon}_i$ is a vector describing the deviation of object $i$ from the features of cluster $s$ and includes measurement errors, while $\bar{\eta_{s_i}}$ describes cluster-specific features. $g_s$ is a loading factor that emphasises the similarity or difference between objects in cluster $s$. In this research the data set refers to a set of the objects, denoting $N$ assets or stocks, and their features are prices across $D$ days in the data set. The variable $i$ is indexing stocks or assets, whilst $d$ is indexing days.

If $g_s = 1$, all objects with $s_i = s$ are identical, whilst if $g_s = 0$, all objects are different. The range of the cluster index is from 1 to $N$ in order to allow for singleton clusters of one object or asset each. 

If one takes Equation 2 as a statistical hypothesis and assumes that both $\bar{\eta}_{s_i}$ and $\bar{\epsilon}_s$ are Gaussian vectors with zero mean and unit variance, for values of $i,s = 1,...,N$, it is possible to compute the probability density $P\left(\lbrace\bar{x_i}\rbrace|\mathcal{G},\mathcal{S}\right)$ for any given set of parameters $(\mathcal{G},\mathcal{S}) = (\lbrace g_s\rbrace,\lbrace s_i\rbrace)$ by observing the data set $\lbrace x_i\rbrace, i,s = 1,...,N$ as a realisation of the common component of Equation 2 as follows \cite{giada2002}:
\begin{equation}
	P\left(\lbrace\bar{x_i}\rbrace|\mathcal{G},\mathcal{S}\right) = \prod\limits_{d=1}^{D}\left\langle \prod\limits_{i=1}^{N}\delta\left(x_i(t) - g_{s_i} \bar{\eta}_{s_i} + \sqrt{1-g^2_{s_i}} \bar{\epsilon}_i\right)\right\rangle.
\end{equation}
The variable $\delta$ is the Dirac delta function and $\langle ... \rangle$ denotes the mathematical expectation. For a given cluster structure $S$, the likelihood is maximal when the parameter $g_s$ takes the values
\begin{equation}
	g^*_s = \begin{cases}
	\sqrt{\frac{c_s - n_s}{n^2_s - n_s}} & \text{for $n_s > 1$},\\
	0 & \text{for $n_s \le 1$}.
	\end{cases}
\end{equation}
The quantity $n_s$ in Equation 4 denotes the number of objects in cluster $s$, i.e.
\begin{equation}
	n_s = \sum\limits_{i=1}^{N}\delta_{s_i,s}.
\end{equation}
The variable $c_s$ is the internal correlation of the $s^{th}$ cluster, denoted by the following equation:
\begin{equation}
	c_s = \sum\limits_{i=1}^{N} \sum\limits_{j=1}^{N} C_{i,j}\delta_{s_i,s}\delta_{s_j,s}.
\end{equation}

The variable $C_{i,j}$ is the \textit{Pearson correlation coefficient} of the data, denoted by the following equation:
\begin{equation}
	C_{i,j} = \frac{\bar{x_i}\bar{x_j}}{\sqrt{\norm{\bar{x_i}^2}\norm{\bar{x_j}^2}}}.
\end{equation}

The maximum likelihood of structure $\mathcal{S}$ can be written as $P\left(\mathcal{G}^*,\mathcal{S}|\bar{x_i}\right) \propto \exp^{D \mathcal{L}(\mathcal{S})}$ (see \cite{S2000}), where the resulting likelihood function per feature $\mathcal{L}_c$ is denoted by
\begin{equation}
	\mathcal{L}_c(\mathcal{S}) = \frac{1}{2} \sum\limits_{s:n_s > 1}^{}\left(\log\frac{n_s}{c_s} + (n_s - 1) \log\frac{n^2_s - n_s}{n^2_s - c_s}\right).
\end{equation}

From Equation 8, it follows that $\mathcal{L}_c = 0$ for clusters of objects that are uncorrelated, i.e. where $g^*_S = 0$ or $c_s = n_s$ or when the objects are grouped in singleton clusters for all the cluster indexes ($n_s = 1$). Equation 8 illustrates that the resulting maximum likelihood function for $\mathcal{S}$ depends on the \textit{Pearson correlation cofficient} $C_{i,j}$ and hence exhibits the following advantages in comparison to conventional clustering methods: 
\begin{itemize}
	\item It is \textbf{unsupervised}: The optimal number of clusters is unknown \textit{a priori} and not fixed at the beginning
	\item The interpretation of results is \textbf{transparent} in terms of the model, namely Equation 2.
\end{itemize}
Giada and Marsili state that $\max_s \mathcal{L}_c(\mathcal{S})$ provides a measure of structure inherent in the cluster configuration represented by the set $\mathcal{S} = \lbrace s_1,...,s_n \rbrace$ \cite{giada2002}. The higher the value, the more pronounced the structure.

\section{Parallel Genetic Algorithms}
In order to localise clusters of normalised stock returns in financial data, Giada and Marsili made use of a \textit{simulated annealing} algorithm \cite{giada2001,giada2002}, with $-\mathcal{L}_c$ as the cost function for their application of the log-likelihood function on real-world data sets to substantiate their approach. This was then compared to other clustering algorithms, such as \textit{K-means}, \textit{single linkage}, \textit{centroid linkage}, \textit{average linkage}, \textit{merging} and \textit{deterministic maximisation} \cite{giada2002}. The technique was successfully applied to South African financial data by Mbambiso et al., using a serial implementation of a \textit{simulated annealing} algorithm (see \cite{mbambiso} and \cite{gebbie}).

\textit{Simulated annealing} and \textit{deterministic maximisation} provided acceptable approximations to the maximum likelihood structure, but were inherently computationally expensive. We promote the use of PGAs as a viable an approach to approximate the maximum likelihood structure. $\mathcal{L}_c$ will be used as the fitness function and a PGA algorithm will be used to find the maximum for $\mathcal{L}_c$, in order to efficiently isolate clusters in correlated financial data.

\subsection{GA principle and genetic operators}
One of the key advantages of GAs is that they are conceptually simple. The core algorithm can be summarised into the following steps: \textit{initialise population}, \textit{evolve individuals}, \textit{evaluate fitness}, \textit{select individuals to survive to the next generation}. GAs exhibit the trait of broad applicability \cite{sivanandam}, as they can be applied to any problem whose solution domain can be quantified by a function which needs to be optimised.

Specific genetic operators are applied to the parents, in the process of reproduction, which then give rise to offspring. The genetic operators can be classified as follows:\newline

\underline{\textit{Selection:}} The purpose of selection is to isolate fitter individuals in the population and allow them to propogate in order to give rise to new offspring with higher fitness values. We implemented the \textit{stochastic universal sampling selection operator}, where individuals are mapped to contiguous segments on a line in proportion to their fitness values \cite{baker}. Individuals are then selected by sampling the line at uniformly spaced intervals. While fitter individuals have a higher probability of being selected, this technique improves the chances that weaker individuals will be selected, allowing diversity to enter the population and reducing the probability of convergence to a local optimum.

\underline{\textit{Crossover:}} Crossover is the process of mating two individuals, with the expectation that they can produce a fitter offspring \cite{sivanandam}. The crossover genetic operation involves the selection of random loci to mark a cross site within the two parent chromosomes, copying the genes to the offspring. A bespoke \textit{knowledge-based crossover} operator was developed for our implementation \cite{cieslakiewicz}, in order to incorporate domain knowledge and improve the rate of convergence.

\underline{\textit{Mutation:}} Mutation is the key driver of diversity in the candidate solution set or search space \cite{sivanandam}. It is usually applied after crossover and aims to ensure that genetic information is randomly distributed, preventing the algorithm from being trapped in local minima. It introduces new genetic structures in the population by randomly modifying some of its building blocks and enables the algorithm to traverse the search space globally.

\underline{\textit{Elitism:}} Coley states that fitness-proportional selection does not necessarily favour the selection of any particular individual, even if it is the fittest \cite{coley}. Thus the fittest individuals may not survive an evolutionary cycle. Elitism is the process of preserving the fittest individuals by inherent promotion to the next generation, without undergoing any of the genetic transformations of crossover or mutation \cite{sivanandam}.

\underline{\textit{Replacement:}} Replacement is the last stage of any evolution cycle, where the algorithm needs to replace old members of the current population with new members \cite{sivanandam}. This mechanism ensures that the population size remains constant, while the weakest individuals in each generation are dropped.\\

Although GAs are very effective for solving complex problems, this positive trait can unfortunately be offset by long execution times, due to the traversal of the search space. GAs lend themselves to parallelisation, provided the fitness values can be determined independently for each of the candidate solutions. While a number of schemes have been proposed in the literature to achieve this parallelisation (see \cite{ismail}, \cite{sivanandam} and \cite{pospichal}), we have chosen to implement the \textit{master-slave} model.

\subsection{Master-slave parallelisation}
Master-slave GAs, also denoted as Global PGAs, involve a single population, but distributed amongst multiple processing units for determination of fitness values and the consequent application of genetic operators. They allow for computation on shared-memory processing entities or any type of distributed system topology, for example grid computing \cite{pospichal}. 
%

Ismail provides a summary of the key features of the master-slave PGA \cite{ismail}: The algorithm uses a single population (stored by the master) and the fitness evaluation of all of the individuals is performed in parallel (by the slaves). Communication occurs only as each slave receives the individual (or subset of individuals) to evaluate and when the slaves return the fitness values, sometimes after mutation has been applied with the given probability. The particular algorithm we implemented is \textit{synchronous}, i.e. the master waits until it has received the fitness values for all individuals in the population before proceeding with selection and mutation. The \textit{synchronous} master-slave PGA thus has the same properties as a conventional GA, except evaluation of the fitness of the population is achieved at a faster rate. The algorithm is relatively easy to implement and a significant speedup can be expected if the communications cost does not dominate the computation cost. The whole process has to wait for the slowest processor to finish its fitness evaluations until the selection operator can be applied.

A number of authors have used the Message Parsing Interface (MPI) paradigm to implement a master-slave PGA. Digalakis and Margaritis implement a synchronous MPI PGA and shared-memory PGA, whereby fitness computations are parallelised and other genetic operators are applied by the master node only \cite{DM2003}. They demonstrate a computation speed-up which scales linearly with the number of processors for large population sizes. Zhang et al. use a centralised control island model to concurrently apply genetic operators to sub-groups, with a bespoke migration strategy using elite individuals from sub-groups \cite{ZLL2012}. Nan et al. used the MATLAB parallel computing and distributed computing toolboxes to develop a master-slave PGA \cite{NPYW2010}, demonstrating its efficacy on the image registration problem when using a cluster computing configuration. 

For our implementation, we made use of the Nvidia CUDA platform to achieve massive parallelism by utilising the Graphical Processing Unit (GPU) Streaming Multiprocessors (SM) as slaves, and the CPU as master.

\vspace{-2mm}
\section{Computational Platform and Implementation}
Compute Unified Device Architecture (CUDA) is Nvidia’s platform for massively parallel high performance computing on the Nvidia GPUs. Compute Unified Device Architecture (CUDA) is Nvidia’s platform for massively parallel high-performance computing on the Nvidia GPUs. At its core are three key abstractions: a hierarchy of thread groups, shared memories, and barrier synchronisation. Full details on the execution environment, thread hierarchy, memory hierarchy and thread synchronisation schemes have been omitted here, but we refer the reader to Nvidia technical documentation \cite{nvidia, nvidia2} for a comprehensive discussion.

\subsection{Specific computational environment}
The CUDA algorithm and the respective testing tools were developed using Microsoft Visual Studio 2012 Professional, with the Nvidia Nsight extension for CUDA-C projects. The following
configurations were tested to determine the versatility of the CUDA clustering algorithms on the following architectures:
\begin{table}[h!]\scriptsize	
	\captionsetup{font=scriptsize}	
	\centering
	\begin{tabular}{|p{1.7cm}|p{5cm}|p{1.1cm}|}
		\hline
		\textbf{Environment} & \textbf{Configuration} & \textbf{Framework}\\
		\hline
			& Windows 7 Professional Service Pack 1 (64-bit), & CUDA 5.5\\
		GTX\_CUDA &	Intel Core i7-4770K CPU@3.5 GHz, 32GB RAM, & (parallel)\\
			&	Nvidia GTX Titan Black with 6GB RAM, & \\
			& CC: 3.0, SM: 3.5 & \\
		\hline
			& Windows 7 Professional Service Pack 1 (64-bit), & MATLAB\\
		GTX\_MATLAB &	Intel Core i7-4770K CPU@3.5 GHz, 32GB RAM, & 2013a\\
			&	Nvidia GTX Titan Black with 6GB RAM, & (serial)\\
			& CC: 3.0, SM: 3.5 & \\
		\hline
			& Windows 7 Professional Service Pack 1 (64-bit), & CUDA 5.5\\
		TESLA\_CUDA &	Intel Core i7-X980 CPU@3.33 GHz, 24GB RAM, & (parallel)\\
			&	Nvidia TESLA C2050 with 2.5GB RAM, & \\
			& CC: 2.0, SM: 2.0 & \\
		\hline
			& Windows 7 Professional Service Pack 1 (64-bit), & MATLAB\\
		TESLA\_MATLAB &	Intel Core i7-X980 CPU@3.33 GHz, 24GB RAM, & 2013a\\
			&	Nvidia TESLA C2050 with 2.5GB RAM, & (serial)\\
			& CC: 2.0, SM: 2.0 & \\
		\hline
	\end{tabular}
	\caption{Development, testing and benchmarking environments}
\end{table}\normalsize
We had the opportunity to test two candidate graphics cards for the algorithm implementation: the Nvidia GTX Titan Black and the Nvidia TESLA C2050. Both cards offer double-precision calculations and a comparable number of CUDA cores and TFLOPS (tera floating point operations per second), however the GTX card is significantly cheaper than the TESLA card. The primary reason for this is the use of ECC (error check and correction) memory on the TESLA cards, where extra memory bits are present to detect and fix memory errors \cite{act}. The presence of ECC memory ensures consistency in results generated from the TESLA card, which is critical for rigorous scientific computing. In further investigations, the authors will explore the consistency of the solution quality generated from the GTX card, and whether the resultant error is small enough to justify the cost saving compared to the TESLA card.

\subsection{Implementation}
The following objectives were considered in this research: 1) investigate and tune the behaviour of the PGA implementation using a pre-defined set of 40 simulated stocks featuring 4 distinct disjoint clusters; 2) identify clusters in a real-world dataset, viz. high-frequency price evolutions of stocks; and 3) test the efficiency of the GPU environment.

\subsubsection{Representation}
We used integer-based encoding for the representation of individuals in the genetic algorithm, i.e.
\begin{equation}
	\textit{Individual} = \mathcal{S} = \lbrace s_1, s_2, ..., s_{i-1}, s_i,...,s_N \rbrace
\end{equation}
where $s_i=1,...,K$ and $i=1,...,N$. Here, $s_i$ is the cluster that object $i$ belongs to. In terms of the terminology pertaining to GAs, it means that the $i^{th}$ gene denotes the cluster that the $i^{th}$ object or asset belongs to. The numbers of objects or assets is $N$, thus to permit the possibility of an all-singleton configuration, we let $K=N$. This representation was implemented by Gebbie et al. in their serial GA and was adopted in this research \cite{gebbie}.

\subsubsection{Fitness function}
The Giada and Marsili maximum log-likelihood function $\mathcal{L}_c$, as shown in Equation 8, was used as the fitness function. This is used to determine whether the cluster configuration represents the inherent structure of the data set, i.e. it will be used to detect if the GA converges to the fittest individual, which will represent a cluster configuration of correlated assets or objects in the data set.

\subsubsection{Master-slave PGA implementation}
The unparalellised MATLAB GA implementation of the likelihood function by Gebbie, Wilcox and Mbambiso \cite{gebbie} served as a starting point. In order to maximise the performance of the GA, the application of genetic operators and evaluation of the fitness function were parallelised for the CUDA framework \cite{cieslakiewicz}. A summarised exposition is presented here.

Emphasis was placed on outsourcing as much of the GA execution to the GPU and made use of GPU memory as extensively as possible \cite{zhang}. The master-slave PGA uses a single population, where evaluation of the individuals and successive application of genetic operators are conducted in parallel. The global parallelisation model does not predicate anything about the underlying computer architecture, so it can be implemented efficiently on a shared-memory and distributed-memory model platform \cite{sivanandam}. By delegating these tasks to the GPU and making extensive use of GPU memory, this minimises the data transfers between the host and device. These transfers have a significantly lower bandwidth than data transfers between shared or global memory and the kernel executing on the GPU. The algorithm in \cite{gebbie} was modified to maximise the performance of the master-slave PGA and have a clear distinction between the master node (CPU), which controls the evolutionary process by issuing the commands for the GA operations to be performed by the slave nodes (GPU streaming multiprocessors). The pseudo-code for the algorithm implemented is shown in Algorithm 1.
\newpage
\begin{algorithm}[h!]\small
	\caption{Master-slave PGA for cluster identification}
	\begin{algorithmic}
		\STATE Initialise ecosystem for evolution
		\STATE Size the thread blocks and grid to achieve greatest parallelisation
		\STATE \textbf{ON GPU}: Create initial population
		\WHILE{TRUE}
		\STATE \textbf{ON GPU}: Evaluate fitness of all individuals
		\STATE \textbf{ON GPU}: Evaluate state and statistics 
		\STATE \textbf{ON GPU}: Determine if termination criteria are met
		\IF{YES}
		\STATE Terminate ALGO; Exit While loop;
		\ELSE
		\STATE Continue
		\ENDIF
		\STATE \textbf{ON GPU}: Isolate fittest individuals
		\STATE \textbf{ON GPU}: Apply elitism
		\STATE \textbf{ON GPU}: Apply scaling
		\STATE \textbf{ON GPU}: Apply genetic operator: selection
		\STATE \textbf{ON GPU}: Apply genetic operator: crossover
		\STATE \textbf{ON GPU}: Apply genetic operator: mutation
		\STATE \textbf{ON GPU}: Apply replacement (new generation created)
		\ENDWHILE
		\STATE Report on results
		\STATE Clean-up	(Deallocate memory on GPU/CPU; Release device)
	\end{algorithmic}
\end{algorithm}\normalsize

To achieve data parallelism and make use of the CUDA thread hierarchy, we mapped individual genes onto a 2-dimensional grid. Using the representation shown in Equation 9, assuming a
population of 400 individuals and 18 stocks:
\small\begin{equation*}
	\textit{Individual}_1 = \lbrace 1, 2, 4, 5, 7, ..., 6 \rbrace
	\vspace{-4mm}
\end{equation*}

\begin{equation*}
	\textit{Individual}_2 = \lbrace 9, 2, 1, 1, 1, ..., 2 \rbrace
	\vspace{-3mm}
\end{equation*}

\begin{equation*}
	\textit{Individual}_3 = \lbrace 3, 1, 3, 4, 6, ..., 2 \rbrace
	\vspace{-6mm}
\end{equation*}

\begin{equation*}
	\vdots
	\vspace{-6mm}
\end{equation*}

\begin{equation*}
	\textit{Individual}_{400} = \lbrace 8, 1, 9, 8, 7, ..., 3 \rbrace
\end{equation*}\normalsize
would be mapped to grid cells, as illustrated in Figure 1. The data grid cells are mapped to threads, where each thread executes a kernel processing the data cell at the respective $xy$-coordinate.

\begin{figure}[h!]
\captionsetup{font=scriptsize}	
\centering
\includegraphics[width=3in]{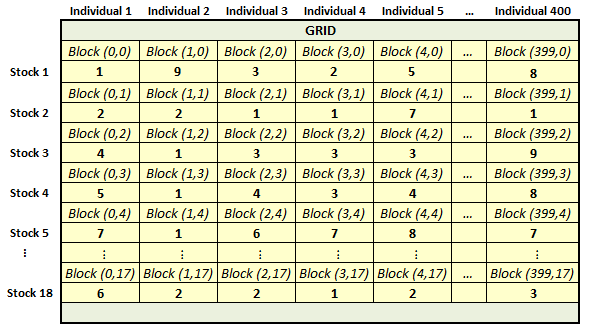}
\caption{Mapping of individuals onto the CUDA thread hierarchy}
\label{fig_mapping}
\vspace{-3mm}
\end{figure}

Given the hardware used in this investigation (see Table 1), Table 2 outlines the restrictions on the permissible stock universe and population sizes imposed by the chosen mapping of individual genes to threads. A thread block dimension of 32 is chosen for larger problems, since this ensures that the permissible population size is larger than the number of stocks to cluster.

\begin{table}\scriptsize
	\captionsetup{font=scriptsize}	
	\centering
	\begin{tabular}{|l|c|c|}
		\hline
		\textbf{Graphics card} & \textbf{Nvidia GTX Titan Black} & \textbf{Nvidia Tesla C2050}\\
		\hline
		\textbf{Compute capability} & 3.5 & 2.0\\
		\textbf{SMs} & 15 & 14\\
		\textbf{Max threads / thread block} & 1024 & 1024\\
		\textbf{Thread block dimension} & 32 & 32\\
		\textbf{Max thread blocks / multiprocessor} & 16 & 8\\
		\textbf{Max number of stocks} & 3840 & 3584\\
		\textbf{Max population size} & 17 472& 18 720\\
		\hline
	\end{tabular}
	\caption{Restrictions on number of stocks and population size. For the Tesla card, $\textit{Max number of stocks} = (14) * (1024/32) * 8 = 3584$ and $\textit{Max population size} = (65535 / (3584/32)) * 32 = 18720$.}
\end{table}\normalsize

We note that the efficiency of the algorithm may be compromised near the physical limits outlined in Table 2, since the CUDA memory hierarchy would force threads to access high-latency global memory banks more often. However, for the particular domain problem we are considering in this paper, the Johannesburg Stock Exchange consists of around 400 listed companies on its main board, which represents an upper limit on the number of stocks of interest for local cluster analysis. This is well within the physical limits of the algorithm, while still providing scope to extend the application to multiple markets.

The details on the full implementation, as well as specific choices regarding initialisation, block sizes and threads per block, are given in \cite{cieslakiewicz}.

\subsubsection{Key implementation challenges}
A key challenge in CUDA programming is adapting to the Single Program Multiple Data (SPMD) paradigm, where multiple instances of a single program use unique offsets to manipulate portions of a block of data \cite{D2001}. This architecture suits data parallelism, whereas task parallelism requires a special effort. In addition, since each warp (group of 32 threads) is executed on a single SPMD processor, divergent threads in a warp can severely impact performance. In order to exploit all processing elements in the multi-processor, a single instruction is used to process data from each thread. However, if one thread needs to execute different instructions due to a conditional divergence, all other threads must effectively wait until the divergent thread re-joins them. Thus, divergence forces sequential thread execution, negating a large benefit provided by SPMD processing.

The CUDA memory hierarchy contains numerous shared memory banks which act as a common data cache for threads in a thread block. In order to achieve full throughput, each thread must access a distinct bank and avoid bank conflicts, which would result in additional memory requests and reduce efficiency. In our implementation, bank conflicts were avoided by using padding, where shared memory is padded with an extra element such that neighbouring elements are stored in different banks \cite{BHS2012}.

CUDA provides a simple and efficient mechanism for thread synchronisation within a thread block via the \textit{\_\_syncthreads()} barrier function, however inter-block communication is not directly supported during the execution of a kernel. Given that the genetic operators can only be applied once the entire population fitness is calculated, it is necessary to synchronise thread blocks assigned to the fitness computation operation. We implemented the CPU implicit synchronisation scheme \cite{XF2010,nvidia3}. Since kernel launches are asynchronous, successive kernel launches are pipelined and thus the executions are implicitly synchronised with the previous launch, with the exception of the first kernel launch. Given the latency incurred on calls between the CPU and GPU, and the consequent drag on performance, GPU synchronisation schemes were explored which achieve the required inter-block communication. In particular, GPU simple synchronisation, GPU tree-based synchronisation and GPU lock-free synchronisation were considered \cite{XF2010}.

Ultimately, the GPU synchronisation schemes were too restrictive for our particular problem, since the number of thread blocks would have an upper bound equal to the number of SMs on the GPU card. If the number of thread blocks is larger than the number of SMs on the card, execution may deadlock. This could be caused by the warp scheduling behaviour of the GPU, whereby active thread blocks resident on a SM may remain in a \textit{busy waiting state}, waiting for unscheduled thread blocks to reach the synchronisation point. While this scheme may be more efficient for smaller problems, we chose the CPU synchronisation scheme in the interest of relative scalability.

\subsubsection{Data pre-processing}
To generate the $N$-stock correlation matrices to demonstrate the viability of the algorithm on real-world test data, data correlations were computed on data where missing data was addressed using zero-order hold interpolation \cite{wilcoxgebbie}. The market mode was removed using the method suggested by Giada and Marsili \cite{giada2001} using a recursive averaging algorithm. A covariance matrix was then computed using an iterative online exponentially-weighted moving average (EWMA) filter with a default forgetting factor of $\lambda=0.98$. The correlation matrix was computed from the covariance matrix and was cleaned using random matrix theory methods. In particular, Gaussian noise effects were reduced by eliminating eigenvalues in the Wishart range in a trace-preserving manner \cite{wilcoxgebbie}. This enhanced the clusters and improved the stability of estimated sequence of correlation matrices.

\subsubsection{Data post-processing}
Computed cluster configurations are read from the CUDA output flat file. Successively, an adjacency matrix is constructed by using data values from the correlation matrix in conjunction with computed cluster configuration of the respective data set. The adjacency matrix is then used to construct a disjoint set of Minimal Spanning Trees (MSTs), each tree capturing the interconnectedness of each cluster. Each MST exhibits $n_s-1$ edges, connecting the $n_s$ stocks of the cluster in such a manner that the sum of the weights of the edges is a minimum. Kruskal's algorithm was used to generate the MSTs, which depict the linkages between highly correlated stocks, providing a graphical visualisation of the resultant set of disjoint clusters \cite{kruskal}.


\vspace{-2mm}
\section{Data and Results}
\subsection{Data}
This investigation used two sets of data: the \textit{training set} and the \textit{test set}. The \textit{training set} consisted of a simulated time series of 40 stocks which exhibit known distinct, disjoint clusters. The recovery of these induced clusters was used to tune the PGA parameters. The \textit{test set} consisted of actual stock quoted midprice ticks aggregated into 3-minute bars from 28 September 2012 to 10 October 2012, viz. approximately 1800 data points for each stock. Stocks chosen represent the 18 most liquid stocks on the JSE for that period, according to traded volumes. For both data sets, correlation matrices were constructed from the time series data, as described in Section 4.2.4, to serve as inputs for the clustering algorithm. The \textit{test set} results below show the summary statistics from a set 1760 correlation matrices of 18 JSE stocks.

\subsection{Results}
We show a sample set of results here. Further discussion regarding aspects of the analysis are given in \cite{cieslakiewicz}.

\subsubsection{Optimal algorithm settings}
Various investigations were undertaken to identify optimal adjoint parameters for the PGA. In each case, the algorithm was successively applied to the \textit{training set}, with known disjoint clusters. Settings were varied until the rate of convergence was maximised. Once the optimal value for each adjoint parameter had been determined from the \textit{training set}, the optimal algorithm configuration was deployed on the \textit{test set}. In further investigations, the authors will study the effect of various adjoint parameter choices on the rate of convergence and algorithm efficiency for varying stock universe sizes.

The following optimal configuration for the PGA was deployed on the \textit{test set}, given a \textit{population size} of 1000:
\begin{table}[h!]\scriptsize
	\captionsetup{font=scriptsize}	
	\centering
	\begin{tabular}{|l|l|}
		\hline
		\textbf{Adjoint parameter} & \textbf{Value}\\
		\hline
		Number of generations &	400\\
		\hline
		Crossover probability ($P_c$) & 0.9\\
		\hline
		Mutation probability ($P_m$) & 0.1\\
		\hline
		Error tolerance & 0.00001\\
		\hline
		Stall generations ($G_{stall}$) & 50\\
		\hline
		Elite size & 10\\
		\hline
		Crossover operator & Knowledge-based operator\\
		\hline
		Mutation operator & Random replacement\\
		\hline
		Knowledge-based crossover probability & 0.9\\
		\hline
	\end{tabular}
	\caption{Development, testing and benchmarking environment}
\end{table}\normalsize

\subsubsection{Benchmark timing results}
Table 4 illustrates the efficiency of the CUDA PGA implementation, compared to the MATLAB serial GA. Direct comparison between the MATLAB serial GA and CUDA PGA may be biased by the fundamental architecture differences of the two platforms. Nevertheless, we immediately observe a significant 10-15 times performance improvement for the test set cluster analysis run. This can be attributed to the utilisation of a parallel computation platform, a novel genetic operator and the algorithm tuning techniques employed. On the GTX platform, the CUDA PGA takes 0.80 seconds to identify residual clusters inherent in a single correlation matrix of 18 real-world stocks, demonstrating its potential as a near-real-time risk assessment tool. The outperformance of the GTX card is likely explained by the card's relative faster core speed, memory speed, larger memory and memory bandwidth compared to the TESLA card. Although this may justify the use of the more cost-effective GTX card, it is not clear that this performance differential will persist as the size of the stock universe increases, or whether the GTX card preserves solution quality. We note that the scale of the performance improvement over the serial algorithm is not as important as the absolute result of obtaining sub-second computation time. The CUDA PGA thus serves the objective of near-real-time risk assessment, whereby interesting phenomena from emerging stock cluster behaviour can be identified and acted upon to mitigate adverse scenarios. The scalability of these results should be investigated in further research, in particular the impact of the CUDA memory hierarchy on computation time as global memory accesses increase.

These results assume correlation matrices are readily available as inputs for the cluster analysis algorithm. Further research to investigate computationally efficient correlation estimation for high-frequency data is a separate problem in the objective of developing a robust and practical near-real-time risk assessment tool.

Although the results are promising, it is not clear that the SPMD architecture used by CUDA is well-suited for the particular problem considered. The required data dependence across thread blocks restricts the assignment of population genes to threads and results in a large number of synchronisation calls to ensure consistency of each generation. An MPI island model with distributed fitness computation and controlled migration is perhaps a more well-posed solution \cite{WRH1999}, however the cost of the setup required to achieve the equivalent speed-up provided by CUDA is important to consider. This should be explored in further research.

\begin{table*}[t]
	\captionsetup{font=scriptsize}	
	\centering
	\begin{tabular}{|l|l|l|c|c|c|}
		\hline
		\textbf{Environment} & \textbf{Framework} & \textbf{Benchmark} & \textbf{Median Time (s)} & \textbf{Min Time (s)} & \textbf{Max Time (s)}\\
		\hline
		GTX\_CUDA	& CUDA 5.5 & 18-stock test set (optimal config) & \textbf{0.80} & 0.73 & 3.17\\
		\hline
		GTX\_MATLAB	& Serial & 18-stock test set (optimal config) & 7.77 & 6.72 & 13.27\\
		\hline
		TESLA\_CUDA	& CUDA 5.5 & 18-stock test set (optimal config) & 1.39 & 1.36 & 5.51\\
		\hline
		TESLA\_MATLAB	& Serial & 18-stock test set (optimal config) & 15.91 & 13.41 & 26.22\\
		\hline
	\end{tabular}
	\caption{Benchmark computational speed results}
\end{table*}\normalsize

\subsubsection{Interpretation of real world test set results}
In this section, we illustrate a sample of the resultant cluster configurations which were generated from our model, represented graphically as MSTs \cite{mbambiso,cieslakiewicz}. This serves as a particular domain application which provides an example of resulting cluster configurations which have meaningful interpretations. The thickness of the vertices connecting nodes gives an indication of the strength of the correlation between stocks.

The South African equity market is often characterised by diverging behaviour between financial/industrial stocks and resource stocks and strong coupling with global market trends. 

\begin{figure}[h!]
\captionsetup{font=scriptsize}	
\centering
\includegraphics[width=2.8in]{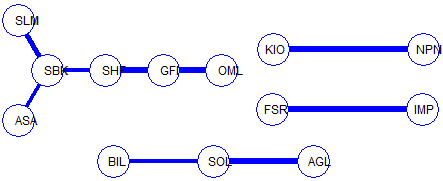}
\caption{Morning trading residual clusters (28 September 2012 09:03)}
\label{fig_cluster1}
\vspace{-3mm}
\end{figure}

In Figure 2, we see 4 distinct clusters emerge as a result of the early morning trading patterns, just after market open. Most notably, a 6-node financial/industrial cluster (SLM, SBK, ASA, SHF, GFI, OML) and a 3-node resource cluster (BIL, SOL, AGL). At face value, these configurations would be expected, however we notice that GFI, a gold mining company, appears in the financial cluster and FSR, a banking company, does not appear in the financial cluster. These are examples of short-term decoupling behaviour of individual stocks due to idiosyncratic factors.

\begin{figure}[h!]
\captionsetup{font=scriptsize}	
\centering
\includegraphics[width=2in]{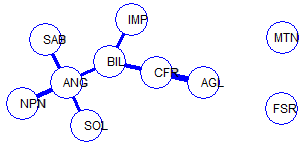}
\caption{Morning trading (after UK open) residual clusters (28 September 2012 10:21)}
\label{fig_cluster2}
\vspace{-3mm}
\end{figure}

Figure 3 illustrates the effect of the UK market open on local trading patterns. We see a clear emergence of a single large cluster, indicating that trading activity by UK investors has a significant impact on the local market. When examining the large single cluster, all of the stocks have either primary of secondary listings in the US and UK. In particular, SAB and ANG have secondary listings on the London Stock Exchange (LSE), whereas BIL and AGL have primary listings on the LSE \cite{jse}. It is also unusual to see such a strong link (correlation) between AGL, a mining company, and CFR, a luxury goods company. This may be evidence that significant UK trading in these 2 stocks can cause a short-term elevated correlation, which may not be meaningful or sustainable.

\begin{figure}[h!]
\captionsetup{font=scriptsize}	
\centering
\includegraphics[width=2.8in]{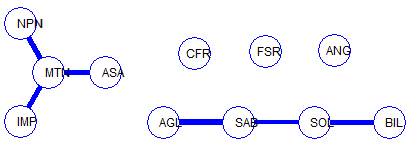}
\caption{Midday trading residual clusters (28 September 2012 12:21)}
\label{fig_cluster3}
\vspace{-3mm}
\end{figure}

Figure 4 considers midday trading patterns. We see that the clustering effect from UK trading has dissipated and multiple disjoint clusters have emerged. CFR has decoupled from AGL in the 2 hours after the UK market open, as we might expect. We see a 4-node financial/industrial cluster (NPN, MTN, ASA, IMP) and 4-node resource cluster (AGL, SAB, SOL, BIL); IMP, a mining company, appears in the financial/industrial cluster.

\begin{figure}[h!]
\captionsetup{font=scriptsize}	
\centering
\includegraphics[width=2.3in]{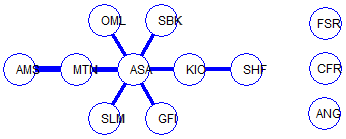}
\caption{Afternoon trading (after US open) residual clusters (28 September 2012 15:33)}
\label{fig_cluster4}
\vspace{-3mm}
\end{figure}

Figure 5 illustrates the effect of the US market open on local trading patterns. Similar to what we observed in Figure 3, we see the emergence of a large single cluster, driven by elevated short-term correlations amongst constituent stocks. This provides further evidence that significant trading by foreign investors in local stocks can cause a material impact on stock market dynamics.


\section{Conclusion}
This paper verifies that the Giada and Marsili \cite{giada2001} likelihood function is a viable, parallelisable approach for isolating residual clusters in datasets on a GPU platform. Key advantages compared to conventional clustering methods are: 1) the method is unsupervised and 2) the interpretation of results is transparent in terms of the model.

The implementation of the master-slave PGA showed that efficiency depends on various algorithm settings. The type of mutation operator utilised has a significant effect on the algorithm’s efficiency to isolate the optimal solution in the search space, whilst the other adjoint parameter settings primarily impact the convergence rate. According to the benchmark test results, the CUDA PGA implementation runs 10-15 times faster than the serial GA implementation in MATLAB for detecting clusters in 18-stock real world correlation matrices. Specifically, when using the Nvidia GTX Titan Black card, clusters are recovered in sub-second speed, demonstrating the efficiency of the algorithm.

Provided intraday correlation matrices can be estimated from high frequency data, this significantly reduced computation time suggests intraday cluster identification can be practical, for near-real-time risk assessment for financial practitioners.

Detecting cluster anomalies and measuring persistence of effects may provide financial practitioners with useful information to support local trading strategies. From the sample results shown, it is clear that intraday financial market evolution is dynamic, reflecting effects which are both exogenous and endogenous. The ability of the clustering algorithm to capture interpretable and meaningful characteristics of the system dynamics, and the generality of its construction, suggests the method can be successful in other domains.

Further investigations include adjoint parameter tuning and performance scalability for varying stock universe sizes and cluster types, quantifying the variability of solution quality on the GTX architecture as a result of non-ECC memory usage and the investigation of alternative cost-effective parallelisation schemes. Given the SPMD architecture used by CUDA, the required data dependence across thread blocks restricts the assignment of population genes to threads and results in a large number of synchronisation calls to ensure consistency of each generation. An MPI island model with distributed fitness computation and controlled migration is perhaps a more well-posed solution to explore \cite{WRH1999}, however the cost of the setup required to achieve the equivalent speed-up provided by CUDA should be justified.\newline

\section*{Acknowledgment}
This work is based on the research supported in part by the National Research Foundation of South Africa (Grant numbers 87830, 74223 and 70643). The conclusions herein are due to the authors and the NRF accepts no liability in this regard.
\bibliographystyle{ieeetr}

\end{document}